# Investigating oxides by electrochemical projection of the oxygen off-stoichiometry diagram onto a single sample

Alexander Stangl*[1,2], Alexander Schmid[3], Adeel Riaz[1], Martin Krammer[3], Andreas Nenning[3], Fjorelo Buzi[4], Fabrice Wilhelm[5], Francesco Chiabrera[4], Federico Baiutti[4], Albert Tarancon[4,6], Jürgen Fleig[3], Arnaud Badel[2,7], Mónica Burriel[1]

* alexander.stangl@neel.cnrs.fr
[1] Université Grenoble Alpes, CNRS, Grenoble INP, LMGP, 38000 Grenoble, France
[2] Université Grenoble Alpes, CNRS, Grenoble INP, Institut Néel, 38000 Grenoble, France
[3] Institute of Chemical Technologies and Analytics, TU Wien, 1060 Vienna, Austria
[4] Catalonia Institute for Energy Research (IREC), 08930 Barcelona, Spain
[5] European Synchrotron Radiation Facility (ESRF), 38054 Grenoble, France
[6] ICREA, 23 Passeig Lluís Companys, 08010 Barcelona, Spain
[7] Université Grenoble Alpes, CNRS, Grenoble INP, G2ELab – Institut Néel, 38000 Grenoble, France

## Abstract:

The oxygen stoichiometry is an essential key to tune functional properties of advanced oxide materials and thus has motivated numerous studies of the oxygen off-stoichiometry diagram, with the aim to determine and control structural, electronic, ionic, electrochemical and optical properties, as well as thermodynamic quantities such as the oxygen storage capacity, among others. Here, a novel approach is developed, which allows to project a broad range of oxygen chemical potentials onto a single thin film sample with unprecedented control via electrochemical polarization. Therefore, a specifically designed electrochemical cell geometry is deployed, resulting in a well-defined, linear, 1D in-plane oxygen concentration gradient, independent of variations in the materials electrical resistivity, whose endpoints can be flexibly controlled via the external $pO_2$ and applied overpotential. This allows for an unparalleled study of materials properties as a continuous function of the oxygen content using spatially resolved tools (spectroscopic, diffraction, microscopy, local electrical probes, etc.) and thereby greatly reduces experimental efforts while also avoiding sample-to-sample variability, multi-step treatments, sample evolution effects, etc. This work presents the proof-of-concept of in-plane oxygen gradients, based on spatially resolved *ex/in situ* and novel fixed-energy X-ray absorption near edge spectroscopy (XANES), X-ray diffraction, ellipsometry and electrical resistivity measurements in hyper-stoichiometric $La_2NiO_{4+\delta}$ and sub-stoichiometric $(La,Sr)FeO_{3-\delta}$ thin films. It thereby demonstrates the readiness and wide applicability of this innovative approach, which can be highly relevant for fundamental as well as applied research.

<u>Keywords:</u> Functional oxides, Functionally graded material, oxygen chemical potential gradient, spatially resolved techniques, mixed ionic electronic conductors

## 1. Introduction:

Oxides are exceptionally versatile materials with promising potential for numerous technologies, including power applications (superconductors,[1] electrochemical energy storage and conversion systems[2,3]), (photo-)catalysis,[4,5] non-volatile memory devices,[6] medical instrumentation,[7] among many others. One peculiarity of these complex materials is their large flexibility to accommodate different oxygen concentrations.[8,9] This non-stoichiometry, $\delta$, corresponds to crystallographic ionic defects in the form of oxygen vacancies or interstitials for sub- and hyper-stoichiometric systems, respectively, and can exceed more than one oxygen per unit cell,[10] with tremendous effects on the phase, microstructure and functional properties.[11–14] For example, the same material can be an antiferromagnetic Mott insulator, a superconductor or a metallic Fermi liquid accompanied by strong changes in its structural, magnetic, electronic and optical state, simply by varying the oxygen content state of these strongly correlated electron systems.[15–17] On the other hand, the presence of oxygen defects in perovskites originates high ionic diffusivities,[18] which in combination with the often relatively high electronic conductivity of metal oxides, gives rise to a new class of functional materials, so-called mixed ionic electronic conductors (MIECs),[19] with great potential for electrochemical applications, such as highly active electrodes for solid oxide cells.[20] Further, the capability to absorb and release oxygen in significant amounts is of strong interest for the catalysis of chemical reactions[21] and for the operation of novel oxygen ion batteries.[22,23]

Uncovering the relationship between ionic defect concentrations and materials properties is therefore of significant importance to various research communities.[11,24] Strategies of tailoring the ionic defect landscape include annealing under controlled temperature and atmosphere,[25] chemical doping,[26] introduction of strain,[27,28] as well as electrochemical methods.[29] However, to investigate the defect-properties correlation, these studies rely on the consecutive modification of the defect concentration of a single sample (linear, multi-step approach), or several samples exposed to different oxygen chemical potentials and thus different conditions (parallel, multi-sample approach), making it time consuming and subject to sample-to-sample variabilities, material evolution/degradation processes and so forth. Recently, a new approach has emerged based on the electrochemical formation of a continuous defect gradient within a single sample by establishing a locally varying electrochemical overpotential (see section 2.1 for details). This allows to map the property evolution as function of the defect concentration using spatially resolved techniques.[30–33] On top, functionally graded materials (FGM) possess the capability to maximize performance efficiency for different tasks at different locations, while maintaining the benefits of deploying a single material only (*e.g.* mechanical stability, recyclability, avoided reactivity), with the potential to disrupt emerging technologies, such as batteries.[34,35]

Typically, the electrical conductivity changes drastically with varying oxygen stoichiometry in functional oxides[36–39] and co-constitutes the community's interest in these materials, *e.g.* for neuromorphic computing.[40] However, to achieve controlled overpotential profiles following aforementioned approach, the electrical conductivity of the oxide ought to be invariant upon changes in oxygen content, which strongly limits its applicability to a small number of materials and/or reduced overpotential ranges.[32] Further, the functional layer needs to be sufficiently thick to provide good electronic in-plane conductance, while maintaining surface limited exchange reactions (*i.e.,* out-of-plane ionic diffusion must be faster than surface reactions thus demanding a thin layer), tightly narrowing the operational window.

Here, we ultimately overcome these restrictions by introducing an original electrochemical cell design that allows us to produce a highly controlled oxygen chemical potential gradient within a single mixed conducting thin film. Key aspects are a three electrode geometry with additional metallic current leads and a surface capping layer. This oxygen chemical potential and thus oxygen defect gradient can be established in a wide class of versatile and emerging oxide materials, with great flexibility, precision and independent from local, defect-induced changes in electrical conductivity, thus presenting minimized restrictions on film thickness and oxygen exchange activity. This method holds the potential to electrochemically project oxygen chemical potentials ranging from the most reduced to the most oxidised state – and thus the full oxygen off-stoichiometry diagram (or selected regions) – into a single sample, unravel the dependence of functional and structural properties on the oxygen stoichiometry and study the emergence of different phases in an unprecedented manner. In this work, we investigate highly controlled oxygen defect gradients in the hyper-stoichiometric Ruddlesden-Popper phase $La_2NiO_{4+\delta}$ (L2NO4) and the sub-stoichiometric perovskite $La_{0.6}Sr_{0.4}FeO_{3-\delta}$ (LSF) using *ex, in situ* and novel fixed-energy X-ray absorption near edge spectroscopy (XANES), X-ray diffraction, spectroscopic ellipsometry and electrical probes to study chemical, structural, optical and functional properties. The proof-of-concept in these different materials with distinct oxygen defect chemistries demonstrates the wide applicability of this methodology, rendering a novel, highly versatile tool for the correlation of the oxygen defect landscape with functional properties in various MIEC oxides.

## 2. Novel methods:

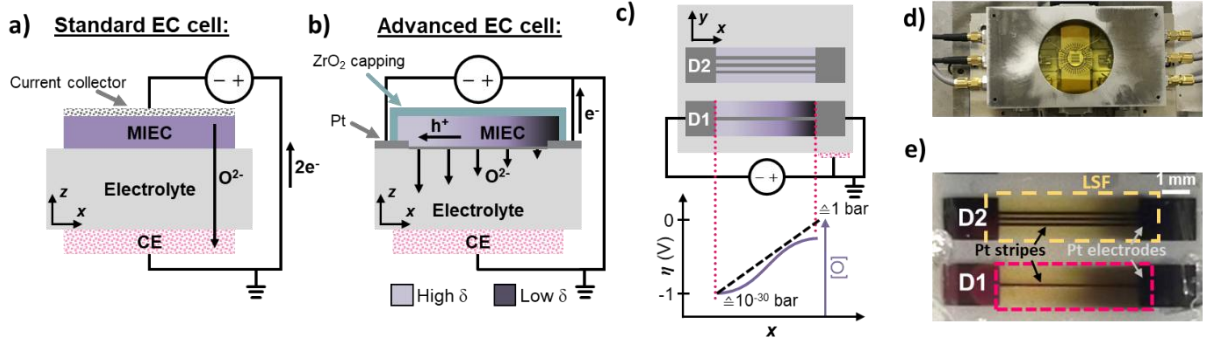

Figure 1: Cross section schematics of (a) standard electrochemical (EC) cell and (b) here developed advanced EC cell architecture. In (b) in-plane electronic (h⁺) and out-of-plane ionic (O²⁻) currents are driven upon polarization of the top left Pt-electrode, while the top right Pt and porous Pt:CGO counter electrode (CE) are grounded. (c) Top view schematic of the micro-fabricated sample with two devices, containing one (D1) or three (D2) Pt stripes between the Pt top electrodes. Perpendicular ionic (out-of-plane, varying) and electronic (in-plane, constant) currents result in a linear variation of the local overpotential (black curve in the bottom graph of (c)), which originates the establishment of a graded in-plane oxygen concentration, [O](x), inside the MIEC (violet curve). Optical images of the in situ setup and the LSF sample are displayed in (d) and (e). The frozen-in oxygen gradient in device D1 and D2 is readily visible due to strong changes of the optical properties of LSF with the oxygen content.

### 2.1. Creating in-plane oxygen gradients

The oxygen stoichiometry of a MIEC is a function of the oxygen chemical potential, $\mu_O$, in the oxide. For oxides being exposed to a gas phase at elevated temperatures, this chemical potential inside the material is often in equilibrium with the oxygen chemical potential in the gas phase, defined by the respective oxygen partial pressure, $pO_2$. For MIEC thin films being used as the working electrode (WE) in a solid oxide electrochemical cell, however, the oxygen chemical potential can also be modified by applying a voltage with respect to a counter electrode (CE). Consequently, oxygen in the MIEC is no longer in equilibrium with the gas phase. In model cells with a thin film WE and a porous CE, the WE surface reaction is typically limiting the kinetics (CE reactions and electrolyte ion conduction are comparatively fast), so that the WE overpotential approximately equals the cell voltage ($\eta \approx U_{WE-CE} \approx U_{\mathrm{appl}}$).[41] This approximation becomes very precise when atmospheric oxygen exchange of the WE is almost entirely blocked by an oxygen impermeable sealing layer, as employed in this study. The applied overpotential $\eta$ then defines the oxygen chemical potential in the MIEC via:

$$\mu_{O,\mathrm{MIEC}} = \mu_{O,\mathrm{CE}}(pO_{2,\mathrm{ref}}) + 2e\eta \qquad \text{Eq. 1}$$

with $\mu_{O,\mathrm{CE}}(pO_{2,\mathrm{ref}})$ denoting the chemical potential of oxygen at the counter electrode (being in equilibrium with the gas phase oxygen partial pressure $pO_{2,\mathrm{ref}}$). The greatly enhanced flexibility provided by non-equilibrium methods (*i.e.* by applying an overpotential) led to strong recent thrust towards electrochemical approaches, with many promising application scenarios.[29,42,43] In standard electrochemical (EC) cells with continuous oxygen pumping, a defined MIEC electrode overpotential, $\eta$, can be established by applying a voltage across a purely ionic conducting electrolyte, sandwiched by two active electrodes, as illustrated in Figure 1(a).[44] Here, $\eta$ alters the non-stoichiometry homogenously within the full MIEC bulk volume (given that the oxygen exchange activity of the MIEC surface is limiting the overall oxygen flow through the cell).

To create an oxygen in-plane gradient in a highly controlled manner, we deploy an advanced electrochemical cell architecture with a three electrode configuration, as schematically drawn in Figure 1(b). It consists of two Pt top electrodes on either side of the MIEC and a fast counter electrode (CE). Applying an external voltage on the top left electrode, while grounding the top right electrode, as well as the counter electrode leads to two current contributions: an in-plane electronic ($h^{\bullet}$) current, $I_{\text{ip}}$, in the oxide thin film (provided that we can neglect the ionic ($O^{2-}$) in-plane contribution of the MIEC due to the dominance of the electronic one) and a purely ionic ($O^{2-}$) current in out-of-plane direction through the YSZ electrolyte, $I_{\text{oop}}(x)$. As the ionic current involves charged oxygen ions, oxygen reactions may take place at the WE and CE surfaces ($1/2\,O_2 \rightleftharpoons O^{2-} + 2h^{\bullet}$), to source or release oxygen species. To establish a well-defined, linear overpotential gradient along $x$, one needs to fulfil the following requirements: (I) the electronic in-plane current is sufficiently higher than the ionic out-of-plane current, $I_{\text{ip}} \gg I_{\text{oop}}$, (II) the total in-plane resistance does not vary locally, (III) the polarization resistances of the ionic conducting substrate and the active counter electrode are much smaller compared to the MIEC and (IV) oxygen incorporation of the MIEC is surface limited, *i.e.* surface exchange reactions of the MIEC are slower than bulk diffusion (with respect to the film thickness). If the conditions (I) and (II) are met, an applied voltage, $U_{\text{appl}}$, at the biased electrode decreases linearly along $x$ reaching zero at the grounded electrode, as expected for an ohmic conductor:

$$U(x) = \left(1 - x/L\right)U_{\text{appl}} \qquad \text{Eq. 2}$$

with the distance from the biased electrode, $x$, and the separation length of the two top electrodes, $L$. This is ensured in our cell configuration by deploying metallic stripes between the two top electrodes, as shown in Figure 1(c), which dominate the electronic transport in in-plane direction. On the other hand, (III) and (IV) guarantee the almost complete translation of the locally sensed voltage into an overpotential ($\eta(x) \cong U(x)$)[*] and a homogenous oxygen profile in the MIEC along $z$, respectively. Both points can be fulfilled via the introduction of an inert capping layer, which strongly reduces the rate of oxygen incorporation at the MIEC surface. The applied voltage thus acts at the same time as linearly varying overpotential, $\eta(x)$, altering the local chemical potential of oxygen, $\mu_{\text{O,MIEC}}(x)$ according to Eq. 1, which results consequently in a well-controlled in-plane oxygen gradient. As indicated schematically in Figure 1(b), the out-of-plane ionic current through the electrolyte varies laterally, and decreases with distance from the biased electrode. For perfectly suppressed oxygen reactions at the MIEC surface, $I_{\text{oop}}$ vanishes in the steady state. Then, the overpotential $\eta$ becomes essentially the local cell voltage of the resulting electrochemical cell with the counter electrode being defined by the oxygen partial pressure and the MIEC electrode exhibiting a laterally varying cell voltage, in accordance with its laterally varying oxygen chemical potential. Actually, the $\mu_{\text{O,MIEC}}(x)$ distribution within the MIEC layer equals that of a Hebb-Wagner polarization experiment (with constant electronic conductivity) with one reversible and one ion-blocking electrode. The effective oxygen partial pressure, $pO_{2,\text{eff}}(x)$, inside the MIEC (assuming ideal gas behaviour), can be expressed using the Nernst equation:[45]

$$pO_{2,\text{eff}}(x) = pO_{2,\text{ref}}\,e^{\frac{4F\eta(x)}{RT}} \qquad \text{Eq. 3}$$

---

[*] For $I_{\text{ip}} \gg I_{\text{oop}}$, the overpotential acting on the MIEC can be written as $\eta(x) = U(x) - I_{\text{oop}}(x) \cdot (R_{\text{YSZ}} + R_{\text{CE}}) \cong U(x)$, if $R_{\text{MIEC}} \gg R_{\text{YSZ}} + R_{\text{CE}}$, with $R$ being the polarization resistances.

with the oxygen pressure inside the electrochemical setup, $pO_{2,\text{ref}}$, and $F$ and $R$ the Faraday and gas constants, respectively. It is important to note that $\mu_{O,\text{MIEC}}$ varies linearly with $x$, which does not imply a linear variation of the oxygen off-stoichiometry. The local oxygen concentration will be modified according to the material specific oxygen non-stoichiometry diagram (also known as Brouwer diagram). This is exemplified with a violet line in the lower graph of Figure 1(c), showing a typical sigmoid shaped curve. We want to highlight that the region of the oxygen chemical potential, projected onto the thin film, can be flexibly selected via the end points, defined by the magnitude of the applied voltage, $U_{\text{appl}}$, and the $pO_{2,\text{ref}}$ in the setup, as demonstrated below.

This innovative approach enables – in combination with spatially resolved techniques – the characterisation of advanced materials properties as a continuous function of the oxygen content in a single sample. It thus renders a unique, efficient and highly useful advancement for the experimentalists' toolbox to correlate functional properties of advanced materials with the oxygen doping state.

### 2.2. Time and spatially resolved X-ray absorption measurements

In this work, we selected XANES as the main characterisation technique, as it provides direct and element specific information on the local oxidation state with high spatial resolution and thus is able to validate our approach. Changes in the oxidation state translate into an energy shift of the element's main edge, *e.g.* to higher energies for an increasing oxidation state. However, XANES scans covering the full edge, and thus an extended energy spectrum, are highly time consuming. For time and spatially resolved XANES measurements, we therefore employed our recently developed methodology, which allows for fast, *in situ* and/or *operando* measurements.[40] In the following we will refer to it as fixed-energy X-ray absorptiometry (FE-XAS). Here, the beam is fixed at a constant energy, close to the K-edge inflection point, while changes in the absorbance intensity, $\Delta\mu(E)$ (not to be confused with the oxygen chemical potential, $\mu_O$), are recorded over time. As the absorption curve rises with the beam energy around the inflection point (positive slope), an increase in measured absorbance intensity corresponds to a shift to lower energies, and vice versa, as comprehensibly illustrated in SI-Figure 1. In a narrow range around the edge energy, the absorbance intensity, $\mu$, varies approximately linearly with constant slope, $k = \left.\frac{d\mu}{dE}\right|_{E_0}$. Energy shifts, $\Delta E$, due to changes in the oxidation state and manifested via changes in the recorded intensity, $\Delta\mu$, can therefore be deduced from a single energy measurement using a rigid shift model, according to:

$$\Delta E = \Delta\mu / k \qquad \text{Eq. 4}$$

with the slope, $k$, obtained from full spectra reference measurements.

The following requirements should be fulfilled to guarantee meaningful data collection. First, the beam source must be sufficiently stable within the timeframe of FE-XAS measurements to not affect the interpretation of intensity variations. Regular beam injections at the ESRF guarantee nearly constant beam currents over long periods and therefore good measurement conditions. Second, changes in oxidation state should be reasonably small to ensure the validity of a rigid shift model, including the preservation of a constant slope around the inflection point within the investigated oxygen stoichiometry range. Third, as the spectrum intensity is also proportional to the probed film volume, variations in thickness can distort and jeopardize this FE-XAS approach. While this can be accounted for to some extend via the subtraction of a reference scan (*e.g.* unpolarised *vs* biased state), sample inhomogeneities should be minimized to allow for high quality spatially resolved measurements.

Fixed energy measurements allow to significantly reduce the acquisition time (*e.g.* from 600 s to 10 s in our case) and therefore provide unparalleled time resolution for X-ray absorption measurements. This enables high throughput and high resolution spatial mappings of the oxidation state, as performed for the first time within this work.

## 3. Results and discussion

### 3.1. Oxygen gradient in $La_2NiO_{4+\delta}$

The hyper-stoichiometry of $La_2NiO_{4+\delta}$ emerges via the formation of oxygen interstitials, $O^{2-}$, in the LaO rock-salt layer,[46] while charge balance is maintained via the mixed valence of Ni, corresponding to the formation of free electron holes.[47] For the stoichiometric compound with $\delta = 0$, the oxidation state of Ni is 2. With increasing $\delta$, Ni partially oxidises from $Ni^{2+}$ to $Ni^{3+}$, resulting in a shift of the Ni absorption edge in the XANES spectrum and therefore allows the quantification of the oxygen off-stoichiometry via $Ni^{\kappa} = Ni^{2+2\delta}$, with the average Ni oxidation state, $\kappa$ (see experimental section for details). The oxygen excess for bulk L2NO4 is well studied using coulometric titration and thermogravimetric analysis, and typically ranges from $\delta = 0$ to $\delta = 0.2$.[48,49] For thin films, the situation remains vague, with only a few recent efforts to understand the oxygen non-stoichiometry.[25,50]

The non-stoichiometry window of L2NO4 thin films was determined by XANES measurements using standard thin film reference samples with different oxygen concentrations: as deposited, annealed for 2 h at 400 °C in a mixture of 5% $H_2$ in Ar (reduced) and in pure $O_2$ (oxidised), respectively. The Ni edges, shown in Figure 2(a), exhibit the same qualitative characteristics, but are subject to a small shift in energy due to changes in the Ni oxidation state, as clearly seen in the inset. This allows us to focus exclusively on the K-edge energy. The inflection points, marked with black arrows, were determined via the maximum of the first derivative, as depicted in SI-Figure 2(a). The observed shift of approximately 0.6 eV between the reduced and the oxidized sample corresponds to a change in the oxygen stoichiometry of $\Delta\delta \approx 0.1$ (using the calibration curve[51] given in SI-Figure 3).

This sets the frame for expectable shifts for the following *in situ* polarization measurements, for which an L2NO4 electrochemical cell was mounted onto a temperature cell inside the beam chamber (Figure 1(e)). Heating the sample did not affect the edge position and only caused a slight decrease in intensity due to thermal expansion (*c.f.* SI-Figure 2(b)).

Prior to *in situ* measurements, we analysed the electronic response of our electrochemical system at 400 °C. The total and out-of-plane currents are shown in SI-Figure 4(a). Even the initial ionic out-of-plane current is three orders of magnitude smaller than the total current, which verifies, that the electronic in-plane current dominates the overall charge flux. Further, the fact that the out-of-plane current decreases quickly to 0 is a good measure of the effectiveness of the deployed surface capping layer to block surface reactions, proving our device architecture viable.

At 400 °C, a bias of -0.8 V was applied to device D2 and XANES spectra were collected along a line between the polarized and grounded top electrodes (see red markers in the schematic of Figure 2(b)). The derivatives of the spectra reveal a clear shift of the edge position with increasing distance from the biased electrode, as shown in Figure 2(b). This spatial variation of the Ni oxidation state corresponds to a continuous increase of the oxygen concentration with $x$, and thus validates our polarization geometry to establish an in-plane oxygen gradient.

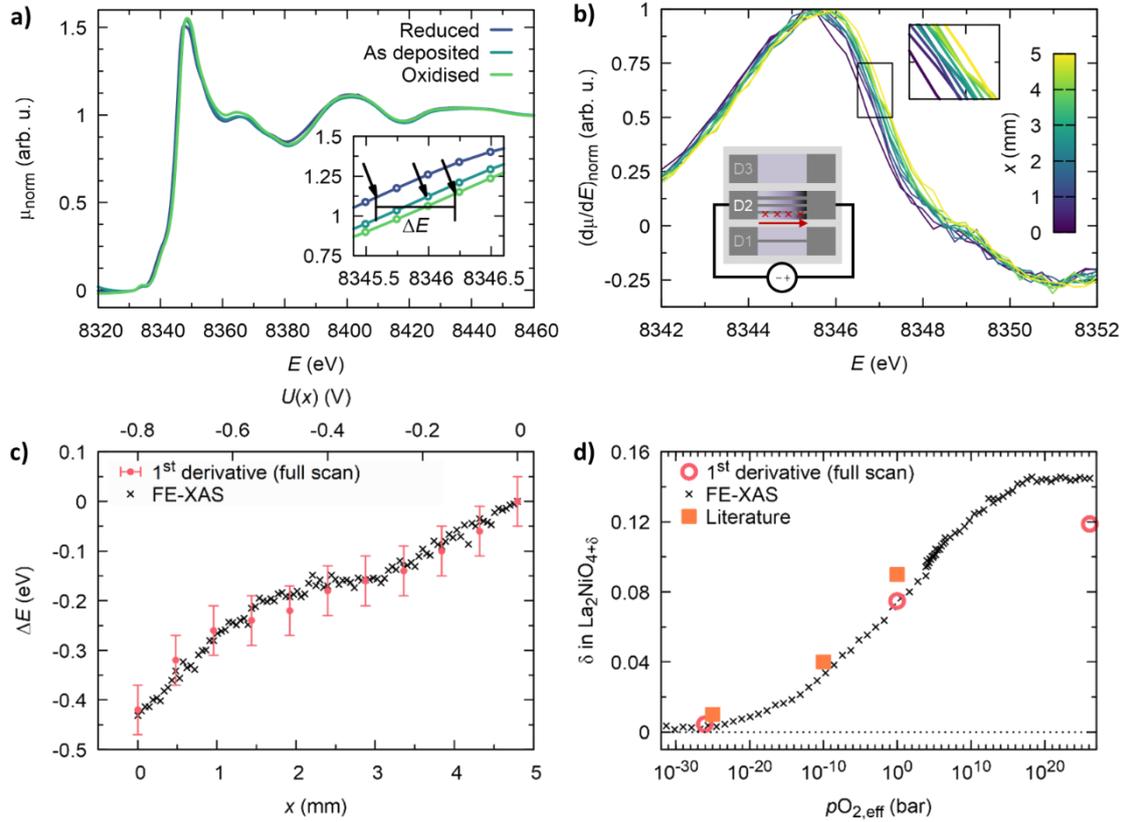

Figure 2: (a) *Ex situ* XANES measurements on L2NO4 reference thin film samples annealed in a mixture of 5 % $H_2$ in Ar (reduced), in $O_2$ (oxidised) and as deposited. The inset shows a magnified region around the inflection point, marked with black arrows. (b) *In situ* measurements at 400 °C in $O_2$ under a bias of -0.8 V as a function of distance from the biased electrode. Red marks in the schematic inset indicate the measurement spots close to the Pt stripes. (c) Spatial oxygen stoichiometry gradient visualized via the Ni K-edge energy shift, obtained from conventional XANES measurements (data shown in (b)) as well as novel high resolution FE-XAS measurements. (d) Calculated oxygen non-stoichiometry of the $La_2NiO_{4+\delta}$ thin film vs. effective oxygen pressure at 500 °C via FE-XAS measurements, compared to values obtained from full scan reference spectra and literature.[25].

The relative energy shift amounts to ≈ 0.45 eV, as depicted as a function of distance in Figure 2(c) (filled circles). Thus, the oxygen excess on the polarized left side of the L2NO4 film is reduced by $\Delta\delta \approx 0.07$, compared to the grounded right side. It is important to note that the observed, approximately linear behaviour is not necessarily expected: the utilized electrode geometry is designed such that the oxygen chemical potential varies linearly (as confirmed below), which does not imply a linear change of the Ni oxidation state. In fact, the shape of the curve is expected to reflect the material specific non-stoichiometry diagram.

To obtain high spatial resolution we deployed FE-XAS, our recently developed fast alternative to conventional XANES measurements.[40] Therefore, the beam energy was fixed to $E_0 = 8345.8$ eV (corresponding approximately to the K-edge energy of the non-biased state) and changes in intensity were recorded at different $x$ positions along a line, as shown in SI-Figure 5(a). Superimposed modulations of the intensity due to changes in film thickness (and thus probed volume) were removed by subtracting a reference scan at zero bias and the corrected absorbance intensity variations were converted into an energy shift using the rigid shift model outlined in section 2.2. The resulting curve is depicted in Figure 2(c). We find very good agreement between the conventional XANES method and our new FE-XAS method: we obtain the same overall energy shift of 0.45 eV, as well as very similar local deviations from linearity. As the shift remains smaller than the value obtained for the reference samples (0.6 eV), the thermodynamic maxima have not been reached. To quantify the full oxygen non-stoichiometry window of L2NO4, *i.e.* as relevant for its applicability in memristive devices[25,52], oxygen storage systems[9] and oxygen ion batteries[22,23], we investigated the energy shift as a function of $U_\text{appl}$ at fixed position close to the biased electrode, using FE-XAS measurements at 500 °C. The voltage was stepwise increased from -1.2 V to +1 V, while changes in intensity were recorded at a constant beam energy, as shown in SI-Figure 5(b) (open circles, left scale). We obtained a sigmoid curve with saturation below -1 V and above +0.7 V. Again, FE-XAS data shows very good agreement with full spectra reference scans (open diamonds, intensity normalised to the acquisition time/point).

The resulting energy shift is shown on the right axis of SI-Figure 5(b) (filled circles). The total shift amounts to approximately 0.9 eV, indicating a large change in the oxygen interstitial concentration. Based on the facts, that the XANES edge energy remained constant between RT and 400 °C and that the Ni K-edge position varies nearly linearly with the Ni oxidation state in the range of interest (*c.f.* SI-Figure 3), the oxygen non-stoichiometry can be calculated. Converting the applied voltage into an effective $pO_{2,\text{eff}}$ using the Nernst equation (Eq. 3), allows to present an extended oxygen non-stoichiometry diagram for $La_2NiO_{4+\delta}$ thin films in Figure 2(d). It is noteworthy, that these very high $pO_{2,\text{eff}}$ values do not correspond to a physical pressure, but it is a measure of the oxygen activity, while in closed nanopores in thin film electrodes, oxygen fugacity values exceeding $10^{12}$ bar have been recently reported.[53] Our data follows a typical S-shaped curve, levelling off at low and high pressure due to saturation of the crystal structure to release or incorporate further oxygen interstitials, characteristic for oxide materials. Further decreasing/increasing the applied potential and thus the effective oxygen pressure would eventually lead to the decomposition of the material and was therefore avoided. Based on these measurements, the accessible oxygen excess window of our L2NO4 thin films at 500 °C can be quantified to about $\Delta\delta \approx 0.14$, in good agreement to literature.[54]

For completeness, we compare the continuous FE-XAS data in Figure 2(d) with $\delta$ values obtained via the 1$^\text{st}$ derivative from *in situ* full spectra reference scans at -1, 0 and +1 V (pink triangles) at 500 °C and *ex situ* literature data (orange squares). Literature values correspond to $La_2NiO_{4+\delta}$ thin films annealed at 500 °C in 6 % H$_2$ in Ar ($\approx 10^{-25}$ bar), Ar ($\approx 10^{-10}$ bar) and pure O$_2$ (1 bar) prior to XANES measurements.[25] Generally, very good agreement is obtained between literature values, *in situ* reference data and the continuous FE-XAS curve. Only under large positive bias, at very high effective pressures, we find a deviation of the FE-XAS approach and the conventional quantification method using the first derivative. This small mismatch can be understood due to changes in $k$ under very high $pO_{2,\text{eff}}$, as shown and discussed in SI-Figure 6.

### 3.2. Oxygen gradient in La$_{0.6}$Sr$_{0.4}$FeO$_{3-\delta}$

In the previous section we have demonstrated, that our advanced, yet simple electrode configuration allows to establish a continuous, lateral oxygen gradient in hyper-stoichiometric L2NO4 thin films. Next, we will show in sub-stoichiometric LSF that this oxygen gradient can be frozen-in and fine-tuned to select a specific region of the oxygen defect diagram.

La$_{0.6}$Sr$_{0.4}$FeO$_{3-\delta}$ is an acceptor doped model system with well-established Brouwer diagram, *i.e.* the oxygen non-stoichiometry can vary within the range of $0 \leq \delta \leq 0.2$, see SI-Figure 7(a), corresponding to dopant charge compensation by either electron holes or oxygen vacancies.[55] The filling of oxygen vacancies upon oxidation is compensated by the formation of localised holes on (or in the vicinity of) Fe cations, which therefore undergo a formal partial valence change from Fe$^{3+}$ to Fe$^{4+}$, which can be observed using XANES.

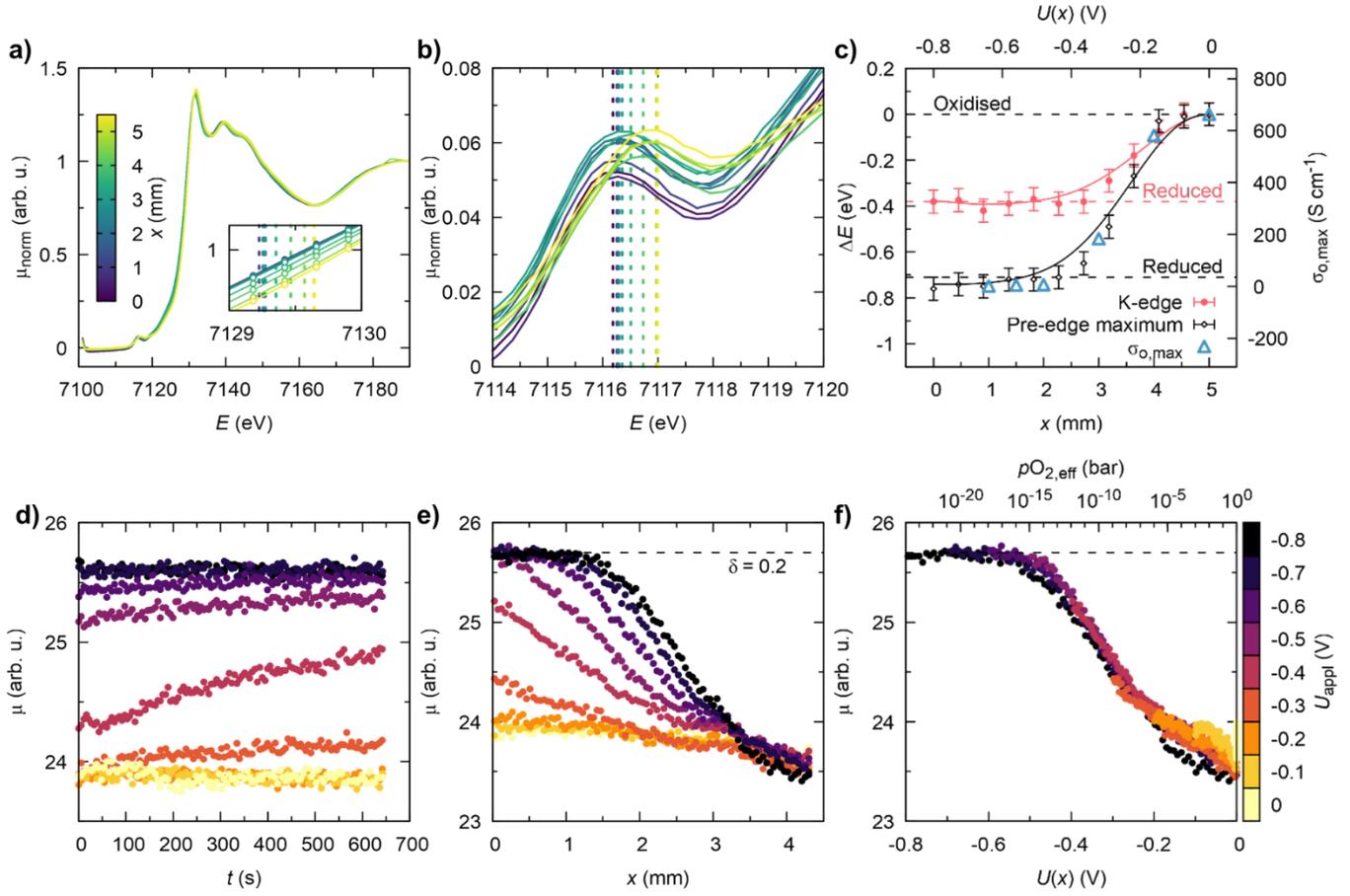

Figure 3: (a-c) Ex situ X-ray absorption spectra recorded along $x$ in LSF sample (D1): (a) Fe K-edge with magnified inflection point and (b) pre-edge maximum. (c) Energy shift of the K-edge and the maximum of the pre-edge as function of distance from the biased electrode, extracted from (a) and (b). Dashed lines correspond to reference samples, solid lines are guides to the eye in (c). Right y axis shows the maximum optical conductivity obtained from spectroscopic ellipsometry at around 1 eV. (d-f) LSF FE-XAS measurements (device D1) at 400 °C: (d) Absorption intensity as a function of time after a change in polarization (step wise increases: 0 → -0.1 → ... → -0.8 V), measured close to the biased electrode ($x = 0$). (e) Intensity line scans between polarized ($x = 0$) and grounded top electrodes under different polarizations. (f) Absorption intensity data from (e) as a function of locally sensed voltage, $U(x) = \left(1 - {x}/{L}\right)U_{appl}$.

An oxygen chemical potential gradient was frozen into LSF by quenching an LSF advanced electrochemical cell from 400 °C to room temperature, while a bias of -0.8 V was maintained across the devices D1 and D2. Due to strong changes in the optical properties of LSF with the oxygen content, the oxygen gradient can be readily seen, as shown in the optical photograph in Figure 1(e) (highlighted with a rectangular for D1 and D2).

*Ex situ* XANES spectra of the Fe K-edge were recorded at different $x$ positions along device D1, shown in Figure 3(a). Two features of these spectra were analysed: the inflection point of the absorption edge (see inset of Figure 3(a)) and the position of the relative maximum at energies below the absorption edge, see Figure 3(b). As expected from the optically visible colour gradient along the device, a clear shift of the raw data with $x$ is obtained, which is caused by local variations of $\delta$.

The position of the pre-edge maximum and the inflection point both shift with distance by approximately -0.7 and -0.4 eV, respectively, relative to the unbiased (grounded) end of the film, as shown in Figure 3(c). Despite the different magnitude of the shift, both features show the same dependence on the position (from right to left): no shift close to the grounded electrode (between 4.5 and 5 mm), followed by a steep decrease between 2.5 and 4 mm and a 2$^{nd}$ plateau for $x < 2$ mm. The relative energies of both plateaus agree well with the energies measured in the reference samples, where the oxygen non-stoichiometry, $\delta$, was set via annealing in O$_2$ and in 5 % H$_2$ in Ar, and is well known for LSF thin films, *i.e.* equal 0 for the oxidised and 0.2 for the reduced sample. This confirms that the oxygen content locally varies between the two stoichiometric plateaus with $\delta \approx 0$ (close to the grounded electrode at $x = 5$ mm) and $\delta \approx 0.2$ (for $x \leq 2$ mm). These endpoints correspond to average iron oxidation states of +3.4 and +3, respectively. The former is actually a mixture of localized electron holes (formally Fe$^{4+}$) and regular Fe$^{3+}$ ions.[55] Specifically, we find values of 7129.85 eV and 7129.48 eV for the inflection point of oxidised and reduced reference samples, and of 7116.73 eV and 7116.00 eV for the pre-edge maximum of oxidised and reduced reference. With these two known oxidation states and their respective absorption energies, we map the absorption energies in Figure 3(c) to a formal iron oxidation state by linear interpolation. The Fe oxidation state thus obtained is shown in SI-Figure 7(b) and follows the expected bulk behaviour. Similar results were obtained for repeated experiments under *in situ* conditions (not shown here). In Figure 3(c), we also report the maximum optical conductivity, $\sigma_{o,\max}$, for the low energy transition (≈1 eV, cf. SI-Figure 8) for different $x$ positions, as retrieved ex-situ by optical ellipsometry mapping. As reported in a previous publication,[56] this spectrum feature is directly proportional to the holes concentration in LSF. The direct superposition between the XANES and optical measurements confirms this relation and validates spectroscopic ellipsometry as a suitable tool for investigating charge carrier concentration in optically active systems.

Notably, the oxygen gradient is stable at room temperature, *i.e.*, the polarization procedure was performed about one week prior to beam time and no changes were observed optically over several months, as expected due to neglectable oxygen diffusion and surface reactions under ambient conditions.

In the following, we demonstrate that this novel approach allows to precisely adjust the span of the oxygen gradient via the magnitude of the applied voltage. Therefore, we performed FE-XAS measurements, recording continuously the intensity at the fixed energy of $E_0 = 7129.7$ eV, corresponding approximately to the absorption edge inflection point of Fe in the non-polarized state. The polarization was stepwise decreased from 0 to -0.8 V with steps of -0.1 V and temporal changes of the intensity (and thus the iron oxidation state) were recorded close to the biased electrode, as shown in Figure 3(d) at 400 °C and in SI-Figure 9 at 450 and 500 °C. The observed increase in intensity with increasing bias corresponds to a shift to lower energies (more reduced state). The curves at different temperatures present the same general features: small polarizations (0 to -0.2 V) barely modify the oxidation state. Similarly, no changes in intensity are observed below -0.6 V, which indicates that all Fe ions are reduced to Fe$^{3+}$ and the oxygen content is not further changed and saturation with $\delta \approx 0.2$ is already reached. In the intermediate regime between -0.2 V and -0.4 V, significant changes in intensity occur. This is in good agreement with the results obtained on the *ex situ* characterized sample, depicted in Figure 3(c) (compare top $x$ axis, obtained using Eq. 2). These *in situ* FE-XAS measurements provide additional kinetic information on the time scale of the oxygen removal processes, including oxygen diffusion in LSF and YSZ, as well as oxygen reduction and evolution reactions at the counter electrode surface, respectively. At 400 °C, saturation is reached within several hundreds of seconds, while at 500 °C notable changes take place only within the first 100 s, followed by a plateau, which corresponds to the new dynamic equilibrium state.

After reaching saturation, line scans were performed along $x$ between the biased and grounded electrodes, as shown in Figure 3(e). At each polarization level, the intensity follows a continuous line between the two electrodes (at $x = 0$ and 5 mm), revealing again a clear gradient in the local oxygen stoichiometry. With increasing magnitude of the applied voltage, the intensity increases at the biased electrode ($x = 0$), up to the point where $\delta \approx 0.2$ is reached. At the grounded electrode, the monitored intensity remains approximately constant. This is the expected result, as the right endpoint is electronically connected to the counter electrode and thus the oxygen chemical potential there is fixed by the $pO_2$ in the atmosphere (1 atm), while the left endpoint is defined via the Nernst equation and thus depends on the applied bias. Due to the special electrode geometry deployed here, the effective oxygen pressure determining the oxygen content inside the MIEC, can be expressed along the full distance, by combining Eq. 3 and Eq. 4:

$$pO_{2,\text{eff}}(x) = pO_{2,\text{ref}} e^{\frac{4F(1-x/L)U_{\text{appl}}}{RT}} \qquad \text{Eq. 5}$$

This exemplifies clearly, how this approach can be used to project different regions of the oxygen chemical potential onto a single thin film sample. For small polarizations, e.g. -0.3 V, a narrow region is selected, with a rather small oxygen gradient along the film width. Increasing the endpoint polarization to about -0.5 V projects the full oxygen non-stoichiometry range of LSF onto the sample. For higher voltages, a constant intensity close to the biased electrode is observed. Once the most reduced state is obtained, no further changes in the Fe oxidation state are expected, within the stability region of the phase diagram. Similarly, one can modify the $pO_{2,\text{eff}}$ and apply positive or negative biases to flexibly select desired oxygen chemical potential regimes to be projected onto a single thin film.

Finally, we validate that the applied voltage decreases linearly with distance and translates into a well-defined electrochemical overpotential. Therefore, we convert the position $x$ into the local voltage, $U(x)$, using Eq. 2 and plot the measured intensities at different $U_{\text{appl}}$ as a function of $U(\text{x})$, as given in Figure 3(f). The excellent overlap of the individual data sets confirms the linear mapping of distance to potential and thus demonstrates that our tailored electrode design allows to establish a well-defined, linear potential gradient along the thin film specimen. Different positions are directly linked to different effective oxygen pressures inside the MIEC (and correspondingly different oxygen chemical potentials), as indicated on the upper *x*-axis. The observed superposition of the different $U_{\text{appl}}$-curves further validates that $U(\text{x}) \approx \eta(x)$, *i.e.* the overpotential contributions of the electrolyte and counter electrode can be neglected.

## 3.3. From chemical to structural and functional properties

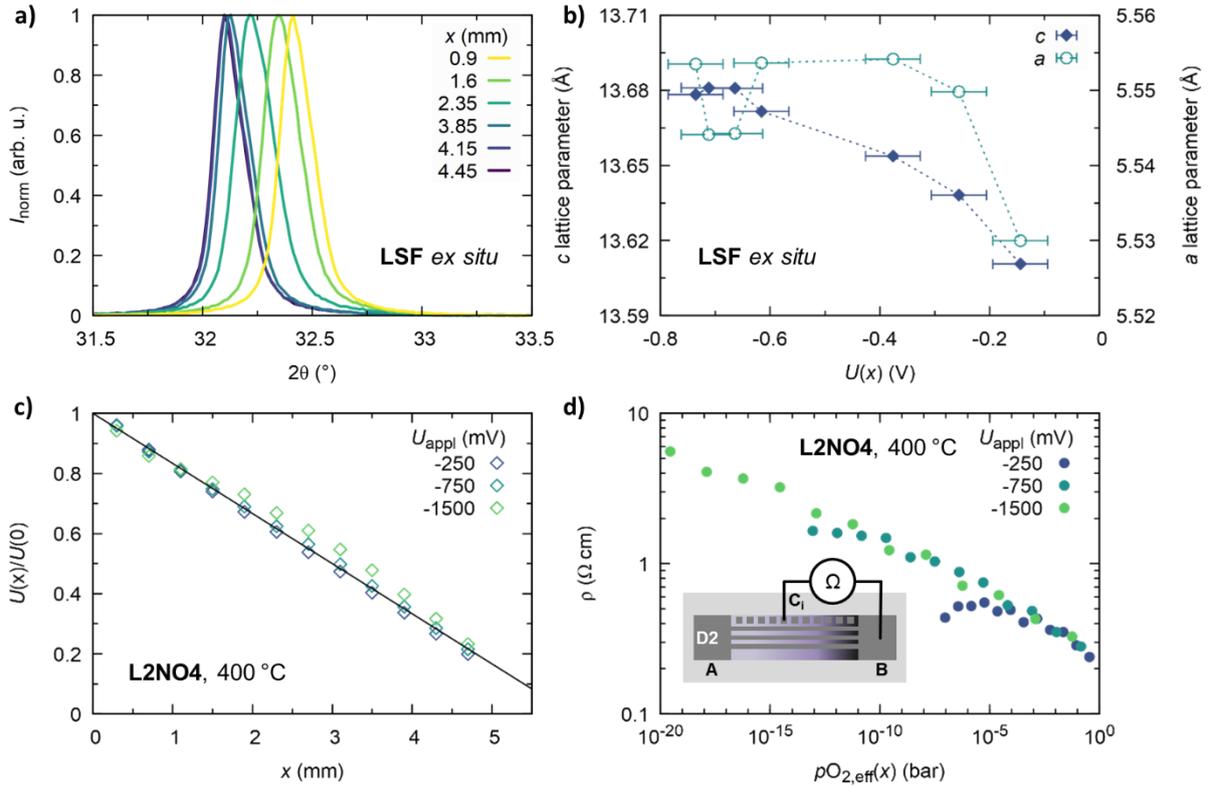

Figure 4: Graded structural and functional properties: (a) ex situ micro-diffraction measurements at different $x$ positions in LSF thin film and (b) resulting crystal lattice parameter. (c) Experimental proof of linear decay of locally sensed voltage at 400 °C for different applied voltages. Black line is a guide to the eye. (d) L2NO4 thin film resistivity gradients as a function of locally sensed effective oxygen partial pressure. Inset sketches the device architecture with additional small electrodes for local electrical measurements.

In the following we demonstrate the translation of the generated chemical gradient into graded structural and functional properties. The chemical expansion of the LSF unit cell was probed using micro X-ray diffraction measurements. Diffraction patterns were recorded at several positions along the frozen-in oxygen gradient, as shown in Figure 4(a), revealing a clear shift to lower angles with increasing oxygen vacancy concentration. The extracted a- and c-axis lattice parameter are given in Figure 4(b) as function of the locally sensed voltage. Such chemical expansion is commonly observed for various types of oxides and can be understood by a combination of two effects, namely an increase of the ionic radius upon reduction of the cation as well as electrostatic repulsion of the positively charged vacancies from the positive cation sub-lattice surrounding it.[57–59] The ability of our herein developed approach to electrochemically induce precisely controlled vacancy gradients provides a promising alternative for the establishment of defect-structure dependencies.

Finally, we turn our attention to the electrical characterisation and precise control of the electronic conductivity in oxides. To resolve spatial changes in resistivity, we fabricated new L2NO4 samples with added Au micro-contacts in between the two top electrodes, as shown in the inset of Figure 4(d). To guarantee good electrical contact, we deposited for these samples the oxide layer first and micro-fabricated the metallic layer on top (*i.e.* in inverse order compared to the samples discussed above), followed by a thin capping layer of $Si_3N_4$.[60]

Defect gradients were established at 400 °C under different biases. Between the application of different polarizations, the sample was cooled to room temperature for electrical characterisation.

Electrical two point measurements were performed between the Au microelectrodes ($C_i$) and the lateral top electrode (B), which corresponds to the orthogonal resistance between the Au stripe and the microelectrode (the resistance between the two top electrodes (A-B) is a factor >$10^3$ smaller). As shown in SI-Figure 10(a), the local resistivity decreases with increasing distance to the polarized electrode (at $x = 0$) and thus increasing oxygen concentration, as expected for L2NO4.[25] Similar results were obtained when measuring the resistance between two adjacent microelectrodes ($C_i$-$C_{i+1}$).

The drastic changes in local resistivity of the MIEC thin film over almost two orders of magnitude may lead to non-linear voltage profiles and therefore not well-defined chemical potential gradients, as it could violate condition (II) (*i.e.* the total in-plane resistance does not vary locally), as defined in section 2.1. We have therefore analysed the local voltage decay during *in situ* polarization measurements. While applying a bias, $U_\mathrm{appl}$, at the top electrode A, the voltage drop between $C_i$-B was probed *in situ* (see sketch in SI-Figure 10(b) for details). Notably, we find a linear decrease of the locally sensed voltage, as shown in Figure 4(c), with only small deviation from linearity even for the highest tested bias of -1500 mV. It is also worth mentioning, that we did not find any potential drop perpendicular to the stripe, as verified in SI-Figure 10(b). Additionally, we measured I-V curves across the devices D1 and D2, containing one and three metallic stripes, respectively, shown in SI-Figure 4(c). The two times higher current across D2 suggests that the in-plane current mainly flows across the metallic stripe(s) and not through the MIEC thin film in agreement with the linear dependence of $U(x)$.

This is remarkable and highlights the major advancement of the novel metallic stripes: despite a change of the materials resistivity over orders of magnitude, the applied voltage still drops only linear across the device length, resulting in a well-defined oxygen chemical potential gradient. Consequently, we can represent the resistivity data from SI-Figure 10(a) as a function of $pO_{2,\mathrm{eff}}(x)$ using Eq. 5 in Figure 4(d). The overlap of the resistivity curves under different polarizations highlights the ability of our methodology to select and project a flexible defect concentration window onto the device, independent of defect dependent electrical properties of the MIEC thin film.

The measurements shown here had all one of the top electrodes at CE potential and thus one side equilibrated with the atmospheric $pO_2$. However, in a 3-electrode device, applying freely chosen voltages to the left and right top electrode enables preparation of single samples with the $pO_2$ window extended above and below atmospheric $pO_2$ as well as samples with a magnified, narrow $pO_2$ range between any two end points within the stability window. We anticipate, that modifications in shape of the current leads even allow for the generation of non-linear overpotential gradients in a well-controlled manner. For example, certain oxygen chemical potential regions can be magnified by increasing the width of a stripe over certain lengths (*i.e.* to reduce the local voltage drop), or having a single stripe narrowing from one side to the other to obtain oxygen activity profiles and thus locally and monotonically varying, complex defect landscapes by design. To account for non-linear scenarios, we can rephrase condition (II) from section 2.1 according to: (II) the local in-plane resistance of the dominant electronic current carrier (*i.e.* the metallic stripe(s)), does not vary upon application of an overpotential.

## 4. Conclusions

We presented a powerful new method to electrochemically establish an oxygen gradient within a single thin film sample. The start and endpoint of this non-stoichiometry gradient can be flexibly tuned, while metallic current leads guarantee a linear decay of the electrochemical overpotential, independent of the MIEC's electronic properties. This device geometry allows the investigation of diverse materials properties as a continuous function of the oxygen content using spatially resolved techniques. The gradient was shown to be stable under bias at elevated temperatures as well as frozen-in at room temperature (and below), thus *operando*, *in situ* and *ex situ* characterisation techniques are viable. This approach therefore provides a unique tool to investigate extended oxygen chemical potential regimes in a single sample in a well-controlled manner, avoiding sample-to-sample variations and time consuming multi-step or multi-sample experiments. In this work we deployed *ex situ*, *in situ* and novel fixed-energy XANES measurements to validate the proposed methodology by studying the oxygen non-stoichiometry in $La_2NiO_{4+\delta}$ and $La_{0.6}Sr_{0.4}FeO_{3-\delta}$ thin films and demonstrated the impact on structural, optical and functional properties. The gradient technique presented here can be readily combined with other spectroscopic, spectrometric, diffraction or local electrical measurements, and thus opens up powerful opportunities for in-depth materials characterization with reduced experimental effort, establish advanced defect-property relationships and ultimately unfold the full materials potential by tailoring defect dependent properties.

# 5. Experimental section

## 5.1. Sample preparation

In this work, two thin film, polycrystalline MIEC materials were investigated. $La_2NiO_{4+\delta}$ (L2NO4) thin films were synthesized using pulsed-injection metal organic chemical vapour deposition (PI-MOCVD)[61] and $La_{0.6}Sr_{0.4}FeO_{3-\delta}$ (LSF) thin films were grown by pulsed laser deposition (PLD).[62,63] Reference samples were deposited directly onto commercial 10×10 mm (100) YSZ single crystal substrates, while for the electrochemical cells, a 100 nm Pt film was micro-fabricated on top of the YSZ substrates using photolithography and metal evaporation prior to the deposition of the oxide layers. The electrochemical cell samples consist of two devices each, as shown in Figure 1(c). The Pt contact pads of device D1 and D2 are connected using 1 and 3 thin Pt stripes (100 μm width, 80 μm spacing for D2), respectively, to ensure sufficient electronic in-plane current with homogeneous lateral resistance. The devices were electronically separated either by mechanical trenching of the oxide layer (L2NO4) or by using shadow masks during MIEC deposition (LSF). The top surface of the MIEC films was capped using a 100 nm $ZrO_2$ layer deposited by PLD to strongly reduce oxygen exchange with the atmosphere. We demonstrated recently that $Al_2O_3$ is a widely available alternative blocking layer.[64] A porous Pt-CGO layer served as fast counter electrode on the unpolished back of the YSZ substrate.[65]

L2NO4 and LSF reference samples were annealed for 2 h at 400 °C and 1 atm in pure $O_2$ (oxidized) and humidified $H_2$ (5%) in Ar (reduced) using a tubular furnace, respectively.

## 5.2. Characterisation

Micro x-ray diffraction measurements were performed on an Empyrean from Panalytical, equipped with an xyz stage. On the primary side the beam was filtered with a focusing mirror to separate the. Cu Kα1,2 radiation, a divergent slit of 1/16° and an 500 μm microfocus collimator. Detection was made with a GalliPix detector. Diffractograms were recorded over a diffraction angle (2 Theta) range of 20 to 110° at several points along the gradient with a spacing of 0.3 mm. The spot size of a single measurement was approximately 300 μm.

Spatially resolved electrical characterisation was performed using a 4-point microprobe station (microworld) equipped with a heating stage (Instec).

The optical properties of the LSF thin films were probed by spectroscopic ellipsometry (UVISEL, Horiba scientific) in a photon energy range from 0.6 to 5.0 eV, with 0.05 eV resolution, and an angle of incident light beam of 70°. The ellipsometry data were modelled and fitted using DeltaPsi2 software (Horiba scientific).

### 5.3. XANES measurements (ex situ and in situ)

In this work, we selected XANES, as it provides direct information on the oxidation state with high spatial resolution. XANES measurements were carried out at the ID12 beamline of the European Synchrotron (ESRF, Grenoble, France). The photon source is the APPLE-II type helical undulator HU-38 at the fundamental harmonic of its emission. Reference samples were measured under vacuum and at room temperature in a top-up 7/8+1 multibunch filling mode for better beam stability. The beam current was nearly constant at 200 mA, with injections every hour. The intensity detector used was a silicon photodiode and the total fluorescence yield (TFY) was collected in backscattered geometry (detector plane at 90° with respect to the beam vector). The cation K-edge energies were determined via the inflection point (maximum of the first derivative) of the XANES spectra. Additionally, for LSF also the position of the pre-edge maximum was analysed. Sporadically occurring spikes (diffraction peaks) in the XANES spectra have been removed.

The *in situ* measurements were performed in oxygen atmosphere (1atmo) using a temperature cell (Nextron), equipped with 6 tips for electrical contacts and a Kapton window, as shown in Figure 1(d). Two silicon drift detectors (SiriusSD) were used to collect the partial fluorescence yield (PFY) in backscattering configuration. A thin (10-20 µm) Mn or Co foil was inserted in front of the SDD to reduce elastic scattering for Fe and Ni K-edge measurements, respectively. A beam spot size of ~200 µm was set with the secondary slits and it was used for low-photon density XANES over large areas. Polyscans were performed in narrow energy ranges around the absorption edge positions of Fe, Ni and Co. For spatially resolved measurements, the beam was focussed using Be refractive lenses. A Keithley 2410 sourcemeter was used for polarization experiments.

For the quantification of the oxidation state of L2NO4, XANES measurements were calibrated using a Ni foil, serving as reference for the Ni (0) oxidation state, with a fixed K-edge position of 8333.0 eV. The resulting offset (2.4 eV) was applied to all L2NO4 XANES measurements. Further, the Ni (II) K edge was determined for a NiO pellet, with good agreement with literature values[51], compare SI-Figure 3. A second order polynomial fit was then used to correlate the measured Ni K-edge energy of the L2NO4 thin films with the average Ni oxidation state, $Ni^\kappa$. Here, we assume formal charge compensation based on the partial oxidation of $Ni^{2+}$ to $Ni^{3+}$ and the formation of doubly charged oxygen interstitials upon oxygen incorporation, *i.e.*: $\frac{1}{2}O_2 + 2Ni_{Ni}^\times \rightarrow O_i'' + 2Ni_{Ni}^\bullet$ (in Kröger-Vink notation), which allows us to determine the oxygen non-stoichiometry via charge balancing: $Ni^\kappa = Ni^{2+2\delta}$, as $\delta = [O_i'']$. Other charge compensation mechanisms, based on singly ionized oxygen interstitials, $O_i'$, or the formation of localised holes on bulk oxygen, $O_O^\bullet$, may take place simultaneously,[66] which would result in larger $\delta$ values. As XANES, however, does not provide information on the oxidation state of oxygen itself, these other mechanisms were not included in our analysis. Self-absorption effects were corrected for in the analysis of the Ni (0) and Ni (II) reference measurements.

# Acknowledgments

This work has received funding from the European Union's Horizon 2020 research and innovation program under grant agreement no. 824072 (Harvestore). This research has benefited from characterization equipment of the Grenoble INP - CMTC platform supported by the Centre of Excellence of Multifunctional Architectured Materials "CEMAM" n°ANR-10-LABX-44- 01 funded by the "Investments for the Future" Program. In addition, this work has been performed with the help of the "Plateforme Technologique Amont" de Grenoble. The authors acknowledge the European Synchrotron Radiation Facility (ESRF) for provision of synchrotron radiation facilities under proposal number MA-5239. We acknowledge Daniel Bourgault (Institut Néel) for access to the high temperature electrical characterisation platform. The X-Ray measurements were carried out by D.I. Werner Artner within the X-Ray Center of the Vienna University of Technology. F.Bu., F.Ba. And A.T. acknowledge the "CERCA Programme / Generalitat de Catalunya".

## Data Availability Statement

The data that support the findings of this study are openly available in zenodo at https://doi.org/10.5281/zenodo.15224271/ and the full beam time data set can be downloaded at https://doi.esrf.fr/10.15151/ESRF-ES-816053550

## Declaration of Competing Interest

The authors declare no known conflict of interest.

## Author Contributions

A.St. suggested the original concept and methodology. A.Sch., A.N., F.Ba., A.T. and M.B. provided additional inputs. A.Sch., A.St., A.R., A.N., F.Ba, F.Bu. prepared the samples. A.St. (as PI), A.Sch., M.K., F.W., A.R., F.Bu. performed the synchrotron measurements. A.St. and A.Sch. analysed the data. M.B., A.T., J.F. and A.B. acquired the funding for research and personnel. A.St. wrote the manuscript. All the co-authors discussed the results and contributed to the revision of the manuscript.

# Supplementary Information:
# Investigating oxides by electrochemical projection of the oxygen off-stoichiometry diagram onto a single sample

Alexander Stangl*[1,2], Alexander Schmid[3], Adeel Riaz[1], Martin Krammer[3], Andreas Nenning[3], Fjorelo Buzzi[4], Fabrice Wilhelm[5], Federico Baiutti[4], Jürgen Fleig[3], Arnaud Badel[2,6], Mónica Burriel[1]

* alexander.stangl@neel.cnrs.fr
[1] Université Grenoble Alpes, CNRS, Grenoble INP, LMGP, 38000 Grenoble, France
[2] Université Grenoble Alpes, CNRS, Grenoble INP, Institut Néel, 38000 Grenoble, France
[3] Institute of Chemical Technologies and Analytics, TU Wien, 1060 Vienna, Austria
[4] Catalonia Institute for Energy Research (IREC), 08930 Barcelona, Spain
[5] European Synchrotron Radiation Facility (ESRF), 38054 Grenoble, France
[6] Université Grenoble Alpes, CNRS, Grenoble INP, G2ELab – Institut Néel, 38000 Grenoble, France

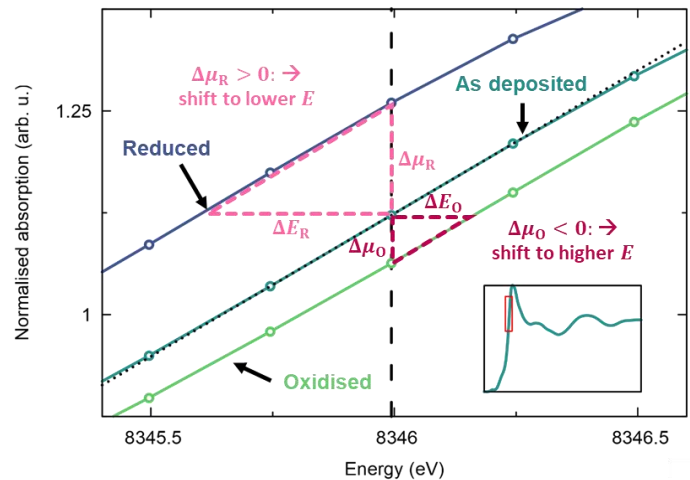

SI-Figure 1: Measurement principle of the fixed-energy X-ray absorptiometry (FE-XAS) technique: the beam energy is fixed close to the edge inflection point (vertical black line) and only changes in intensity, $\Delta\mu$, are recorded. An increase in $\Delta\mu$ corresponds to a shift of the K-edge to lower energies, while a decrease is linked to a higher edge energy (marked with black arrows). In proximity to the inflection point, the slope is approximately constant, as shown by the linear fit for the as deposited state (dotted black line, fitted in the range of 8345.5 to 8346.25 eV). This allows to quantify the energy shift using a rigid shift model, i.e. the shown triangles have the same ratio of $\Delta\mu/\Delta E$, which corresponds to the slope $k = \left.\frac{\partial\mu}{\partial E}\right|_{E_0}$ around the edge energy $E_0$ and thus $\Delta E = \Delta\mu/k$. The shown experimental data corresponds to as deposited, reduced and oxidized $La_2NiO_{4+\delta}$ thin films, see manuscript for details. The inset shows the full edge spectrum, whereas the red rectangular highlights the magnified region of the main figure.

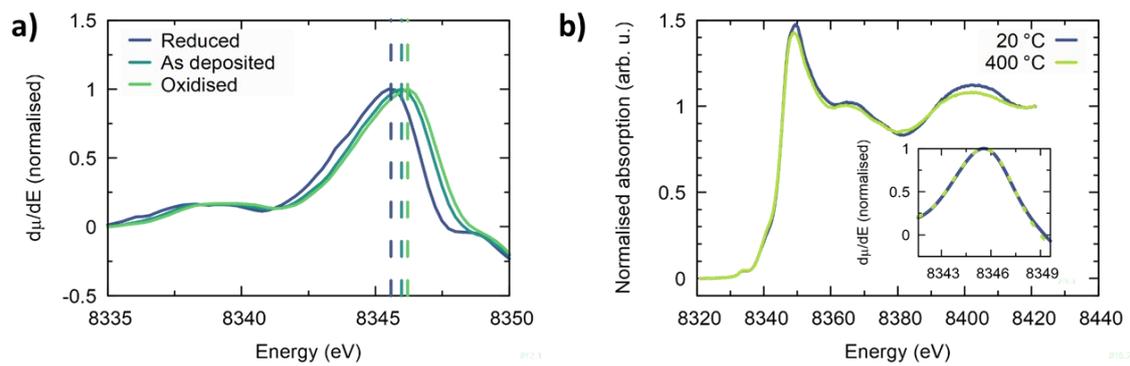

SI-Figure 2: For full scan measurements, the K-edge position is determined via the maximum of the first derivative as exemplarily shown for three L2NO4 references films in (a). In situ XANES spectra at 20 and 400 °C in flowing $O_2$ gas at atmospheric pressure. The edge position remained invariant upon heating, as can be clearly seen from the derivative shown in the inset.

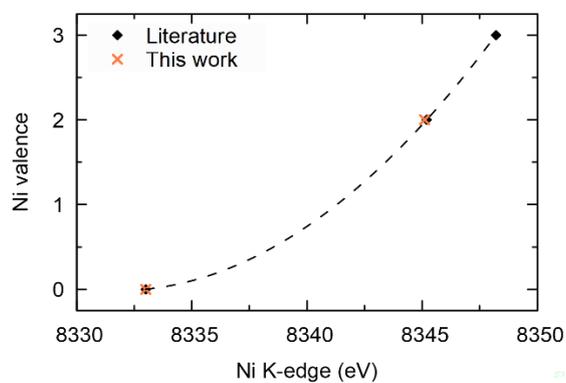

SI-Figure 3: Calibration curve to obtain Ni oxidation state via XANES measurements of the Ni K-edge energy. Literature data points correspond to metallic Ni-foil (Ni (0)) at 8333.0 eV, NiO (Ni (II)) at 8345.2 eV and stoichiometric $LaNiO_3$ (Ni (III)) at 8348.2 eV[51]. Note the almost linear trend between Ni (II) and Ni (III), which allows to approximate changes in oxygen non-stoichiometry, $\Delta\delta$, via the relative K-edge energy shift, $\Delta E$, within the precision of the analysis.

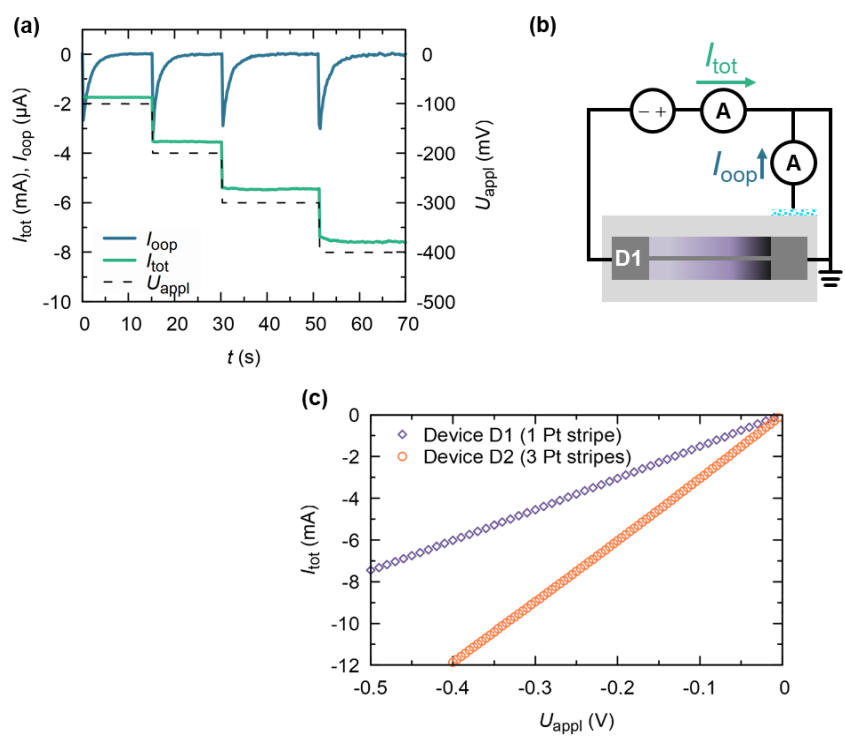

SI-Figure 4: Electrical analysis of an L2NO4 electrochemical cell: (a) Total and out-of-plane currents of device D1 under different polarizations at 400 °C and (b) schematic of the measurement. (c) IV curves measured across device D1 and D2.

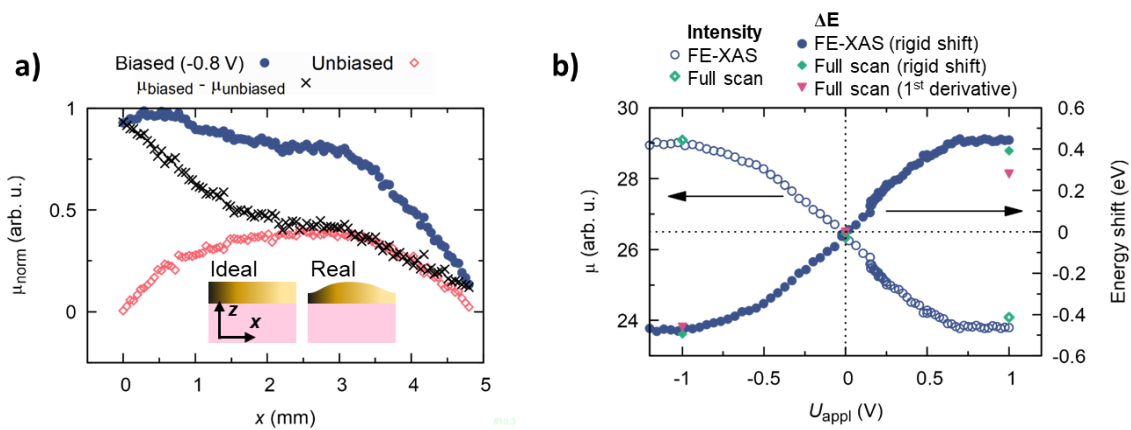

SI-Figure 5: In situ FE-XAS measurements of in L2NO4 (device D2) with fixed beam energy of 8345.8 eV: (a) FE-XAS line scans of unbiased and polarized (-0.8 V) L2NO4 sample. Note, that the measured x-ray absorption intensity at fixed energy is proportional as well to the film thickness, which can be clearly seen for the dome-shaped intensity profile of the unbiased measurement, which we ascribe to an inhomogeneous film thickness, as schematically depicted in the inset. Subtracting the intensity of the unbiased from the biased profile (black markers) allows to minimize the error related to variations in probed film volume and reveals the intensity variations solely due to local changes in the oxidation state. (b) FE-XAS intensity changes as a function of applied voltage, close to the polarized electrode. The right scale shows the energy shift obtained via a rigid shift model. Both, intensity and energy shifts are compared to conventional XANES measurements.

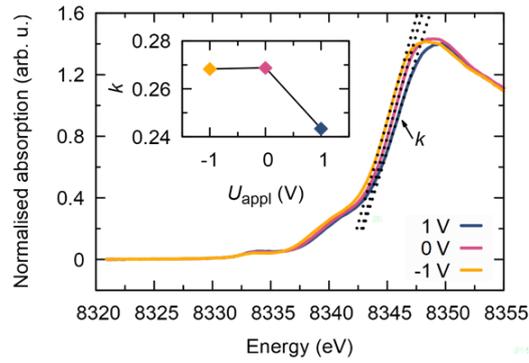

SI-Figure 6: L2NO4 XANES spectra at 500 °C under different polarization (close to the polarized top electrode). Inset shows linear slope around the inflection point for different polarization values. The negative and the unpolarised curves can be matched upon application of an energy shift, with the same slope $k$. On the other hand, for a high positive polarization of +1 V, the shape of the spectrum changes and we find a reduced slope around the inflection point, limiting the precision of FE-XAS measurements to infer the oxidation state.

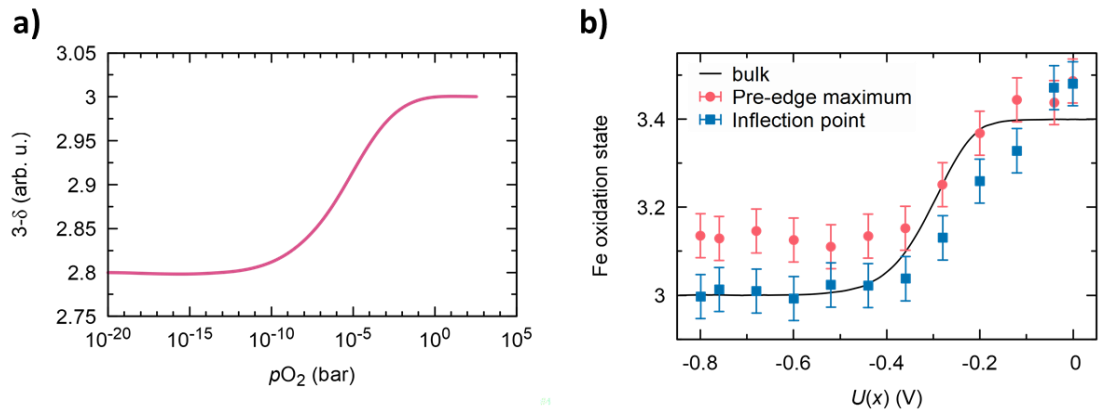

SI-Figure 7: (a) Oxygen non-stoichiometry as function of oxygen partial pressure for $La_{0.6}Sr_{0.4}FeO_{3-\delta}$ bulk at 600 °C, reproduced from literature[55] (b) Fe oxidation state calculated from ex situ XANES data compared to bulk literature data.

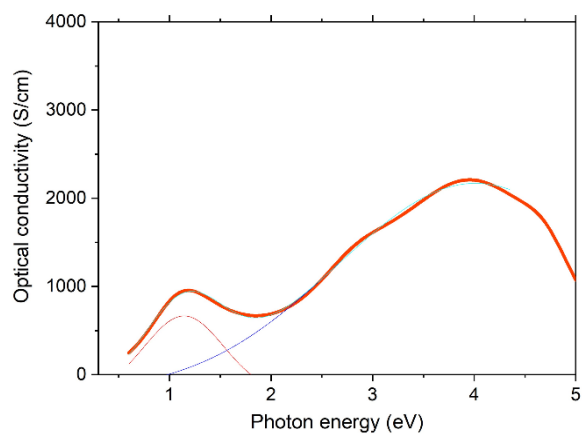

SI-Figure 8: Exemplary ex situ optical spectrum for oxidized LSF (x=5 mm) with fitting curves.

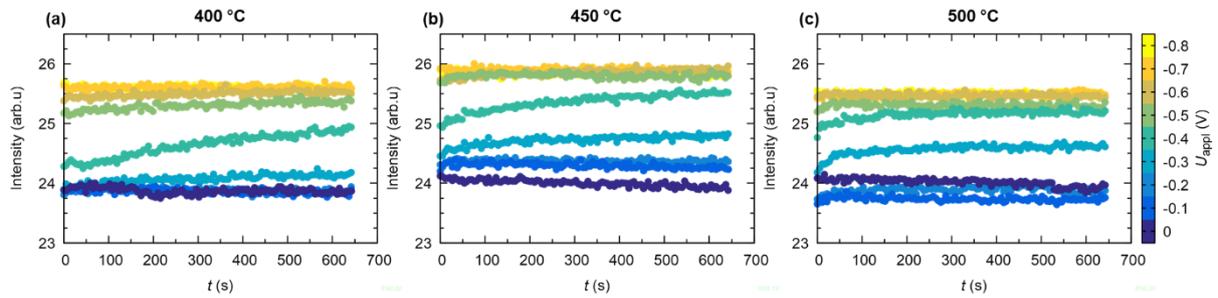

SI-Figure 9: In situ FE-XAS measurements in LSF device: Absorption intensity as a function of time after a change in polarization (0→ -0.1 → -0.2 →…→ -0.8 V), measured close to the biased electrode ($x = 0$) at (a) 400 °C, (b) 450 °C and (c) 500 °C. Changes in intensity are caused by local modification of the Fe oxidation state. With increasing temperature, saturation is reached faster, indicating the thermally activated nature of this process, e.g. oxygen diffusion in the LSF thin film and YSZ bulk as well as surface reactions at the counter electrode.

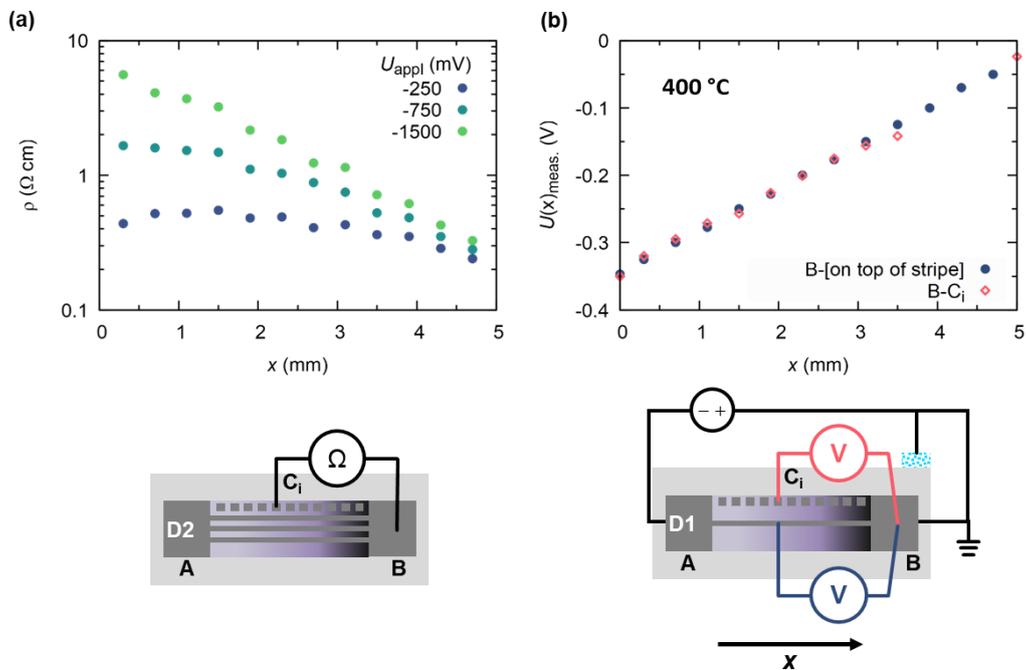

SI-Figure 10: Electrical analysis of an L2NO4 electrochemical cell: (a) Locally sensed voltage (at 400 °C, $U_{appl} = 350$ mV) measured against the ground, B, directly on top of the metallic stripe at different distances (dark blue disks) and on top of small metallic contacts, $C_i$, on the lateral side of the device (pink diamonds), as sketched in the graphic below. (b) Spatially resolved resistivity of L2NO4 thin film after application of different biases at 400 °C, measured at room temperature.